\documentstyle[aps,preprint]{revtex}
\begin{document}
\draft
\title{ Functional Integral approach to quasi-particles in 
Fermi liquid theory }
\author{Tai-Kai Ng}
\address{ department of Physics, Hong Kong University of Science and
           Technology,\\
 Clear Water Bay Road, Kowloon, Hong Kong }

\date{ \today }

\maketitle

\begin{abstract}
   In this paper we propose a new way of organizing many-body perturbation
theory in the Path-integral formulation where a set of 
quasi-particle wave-functionals $\psi$'s are introduced 
and are identified with quasi-particles in Landau
Fermi liquid theory. We show how Fermi liquid theory can be
obtained through $\psi$'s in this new framework of
perturbation theory, where the only assumption is adiabaticity between
non-interacting and interacting states and the quasi-particle
renormalization factor {\em z} does not appear explicitly.
Consequences of our new formulation are discussed.
\end{abstract}


\narrowtext
   For four decades Fermi liquid theory has formed the basis for our
understanding of interacting fermion systems like electron liquids
in metals and normal state of liquid $He^3$. The theory was
originally proposed phenomenologically by Landau\cite{l1,l2}
and was later put on a firm basis by a formal 'proof' using 
diagramatic perturbation theory techniques\cite{l3,b1,no}.  
Although formally exact, the existing perturbation theory 
formulation of Fermi liquid theory has a drawback.
In the original Fermi liquid theory of Landau, the
only assumption was adiabaticity or a one-to-one 
correspondence between non-interacting states (bare particles) and
states when interaction is turned on (quasi-particles).
However, in the existing diagrammatic technique, perturbation
theory was carried out in terms of {\em bare particles}
and the connection between bare particles and
quasi-particles is established by assuming that there is a
non-vanishing overlap $z^{1/2}\sim{O}(1)$ between the two. It would
be much more satisfying if the perturbation theory can be 
formulated directly in terms of quasi-particles where the
introduction of renormalization factor {\em z} can be avoided, 
as is in the original Landau formulation.

   In this paper we shall show that within the path-integral
formulation it is possible to formulate perturbation
theory in terms of a set of quasi-particle
wave-functionals which can be identified with quasi-particles
in Fermi liquid theory. With these wave-functionals Fermi liquid 
theory can be derived directly without referring to the
quasi-particle renormalization factor {\em z}. The
one-particle Green's function will be studied where we shall
show that the quasi-particle renormalization factor $z^{1/2}$
measures the overlap between bare-particle states and our
quasi-particle states on the Fermi surface. For simplicity
we shall consider a gas of spinless fermions
interacting through a scalar potential
$v(r)$. The formulation can be generalized to include spin easily.
We shall restrict us to zero temperature where the
perturbation theory is formulated in terms of real-time Path
Integral and Green's functions. The Hamiltonian is, in momentum space,

\begin{equation}
\label{h}
H=\sum_{\vec{k}}\epsilon(\vec{k})f^+_{\vec{k}}f_{\vec{k}}+\lim_{\eta
\rightarrow0} e^{-\eta|t|}{1\over2}
\sum_{\vec{q}}v(q)\rho(\vec{q})\rho(-\vec{q}),
\end{equation}
where $\epsilon(\vec{k})=(\hbar\vec{k})^2/2m$ and 
$f(f^+)_{\vec{k}}$'s are fermion
annihilation(creation) operators. $\rho(\vec{q})=\sum_{\vec{k}}
f^+_{\vec{k}+\vec{q}}f_{\vec{k}}$ is the density operator for
fermions. Notice that we have inserted a $e^{-\eta|t|}$ factor
in front of the interaction term to emphasize that the states in the
interacting system are obtained from states of the non-interacting
system by turning on the interaction {\em adiabatically}, as is
required in Landau Fermi liquid theory. In particular, the time
evolution operator $U(t,t')$,  when operate on the
ground state of the non-interacting system at $t'=-\infty$, turns
the state into the ground state of the interacting system at time $t$.
$U$ can be expressed in terms of a path integral over (fermionic)
coherent states $\Psi, \Psi^+$\cite{no2}. In particular, 

\begin{mathletters}
\begin{equation}
\lim_{T\rightarrow\infty}Tr\left[U(T,-T)|\right]=\int
D\Psi{D}\Psi^+\int{D}\phi{e}^{{i\over2}\int_{-T}^T
dt'\int{d}^dq{\phi(\vec{q},t')\phi(-\vec{q},t')\over{v}(q)}}
e^{{i\over\hbar}\int_{-T}^T
L(\phi;\Psi^+,\Psi,t')dt'},
\label{u1}
\end{equation}
where
\begin{equation}
\label{l1}
L(\phi;\Psi^+,\Psi,t)=\int{d}^dx\left\{i\hbar\Psi^+(\vec{x},t){\partial\Psi(\vec{x},t)
\over\partial{t}}-{\hbar^2\over2m}|\nabla\Psi(\vec{x},t)|^2
-\lim_{\eta\rightarrow0}e^{-{\eta\over2}|t|}\phi(\vec{x},t)
\Psi^+(\vec{x},t)\Psi(\vec{x},t)\right\},
\end{equation}
\end{mathletters}
is the Lagrangian for a system of {\em non-interacting} fermions moving
in the auxillary field $\phi(\vec{x},t)$, introduced through an
Hubbard-Stratonovich transformation from the interaction term in
the original Hamiltonian. In this form, the
time evolution operator $U$ can be interpreted as
a weighted sum of time evolution operators $U(\phi)$'s, where $U(\phi)$
is the time evolution operator for system of non-interacting
fermions moving in given auxillary field $\phi(\vec{x},t)$, with each
configuration of $U(\phi)$ weighted by the factor $e^{i
\int{d}^dq\int{d}t{\phi(\vec{q})\phi(-\vec{q})\over2v(q)}}$ 
in the sum \cite{no2}. The interacting ground state 
wavefunction can thus be interpreted as superposition
of 'ground state' wavefunctions $|G(\phi)>$'s, generated by
operators $U(\phi)$'s acting on the free fermion ground state at
$t'=-\infty$, and weighted by the same $e^{i\int{\phi^2\over2v}}$ factors.
Notice also that the auxillary field $\phi$ is turned on adiabatically
at $t=-\infty$, as a result of adiabaticity of the original
interacting fermion problem. 

   Next we introduce the Green's function operator $G[\phi]$, which is
a functional of $\phi$ field, with
\begin{mathletters}
\label{gx}
\begin{equation}
G^{-1}[\phi]=i\hbar{\partial\over\partial{t}}+{\hbar^2\over2m}\nabla^2-
\lim_{\eta\rightarrow0}e^{-{\eta\over2}|t|}\phi(\vec{x},t),
\end{equation}
and with Fourier transform
\begin{equation}
\label{gk}
G^{-1}_{k,k'}[\phi]=(\hbar\omega-\epsilon(\vec{k})+i\delta_{\vec{k}})
\delta_{k,k'}-\phi(k-k'),
\end{equation}
\end{mathletters}
where $k=(\vec{k},\omega)$ and $\delta_{\vec{k}}=\delta{sgn}(\xi_{\vec{k}})$,
where $\xi_{\vec{k}}=\epsilon(\vec{k})-\mu$ and $\mu=$ chemical
potential. Notice that the adiabaticity requirement is handled by 
introducing the $i\delta_{\vec{k}}$ term in $G$ as in usual
perturbation theory. The ground state energy, one particle Green's
function, etc, can be determined as appropriate functional averages of
$G$ or functions of $G$ over $\phi$ field\cite{no2}. $G_{k.k'}[\phi]$
can be expanded directly in a power series of $\phi$ as in usual
perturbation theory. However, here we shall
rearrange the perturbation expansion in a slightly different
way, by introducing first the {\em eigenfunctions} and corresponding
{\em eigenvalues} of $G^{-1}[\phi]$, where
$G^{-1}[\phi]\psi_k[\phi]=\lambda_k[\phi]\psi_k[\phi]$. Notice
that both $\psi_k$'s and $\lambda_k$'s are functionals of $\phi$.
The perturbative eigenstates $\psi_k$ satisfies the Lippman-Schwinger
equation
\begin{equation}
\psi_k(x;[\phi])=A_k[\phi]\psi_k^{(0)}(x)+\int{d}^{d+1}xG_0(\lambda_k,
x-x')\phi(x')\psi_k(x';[\phi]),
\label{ls}
\end{equation}
where $x=(\vec{x},t)$, $A_k$ is a renormalization factor determined
by $<\psi_k|\psi_k>=1$ and $G_0(\lambda,x)$ is the Fourier transform of
the unperturbed Green's function, $G_0(\lambda,k')=(\lambda-(\omega'-
\epsilon(\vec{k'})+i\delta_{\vec{k'}})^{-1}$.  
$\psi_k^{(0)}(x)\sim{e}^{-i(\omega{t}-\vec{k}.\vec{x})}$ is an
unperturbed eigenstate of $G_0^{-1}$. The eigenvalue $\lambda_k$
can be written as $\lambda_k=\omega-\epsilon(\vec{k})-\Sigma(k;[\phi])+
i\delta_{\vec{k}}$, where $\Sigma(k;[\phi])
=\int_0^1dg\int{d}^{d+1}x\phi(x)|\psi_k
(x;[g\phi])|^2$,
by the Hellmann-Feynman theorem. $\int_0^1dg$ is a coupling constant
integral. Using the fact that $G^{-1}[\phi]$ is a first order
differential equation in time, it can be shown easily that 
$\psi_{\vec{k},\omega+\Omega}(x;[\phi])=e^{-i\Omega{t}}\psi
_{\vec{k},\omega}(x;[\phi])$ and the self energy
$\Sigma(k;[\phi])=\Sigma(\vec{k};[\phi])$ and renormalization
factor $A_{k}[\phi]=A_{\vec{k}}[\phi]$ are {\em independent} of $\omega$. 
The Green's function $G(x,x';[\phi])$ is given by
\begin{equation}
\label{gs}
G(x,x';[\phi])=\sum_k{\psi_k(x;[\phi])\psi^*_k(x';[\phi])\over\hbar\omega-
\epsilon(\vec{k})-\Sigma(\vec{k};[\phi])+i\delta_{\vec{k}}},
\end{equation}
and the usual perturbation expansion for $G$ can be recovered by
expanding $\psi_k[\phi]$ and $\Sigma(\vec{k};[\phi])$ in power
series of $\phi$. 

    The advantage of introducing eigenstates $\psi_k$'s can be
seen when we expand the Grassman fields $\Psi(x)$ in terms of 
$\psi_k$'s using the completeness relation,
$\Psi(x)=\sum_k\psi_k(x;[\phi])c_k$,
with an analogous relation between $\Psi^+(x)$ and $c^+_k$ fields, where
$c_k$ and $c^+_k$ are Grassman number fields and $\psi_k$'s are
complex-number wavefunctionals of $\phi$.
The time evolution operator $U$ can be expressed in terms of
path integral over $c_k$ and $c^+_k$ fields through an Unitary
transformation. In particular,
\begin{mathletters}
\label{lcc}
\begin{equation}
\label{tru}
\lim_{T\rightarrow\infty}Tr\left[U(T,-T)\right]=
\int{D}cDc^+\int{D}\phi{e}^{i\sum_q
{\phi(q)\phi(-q)\over2v(|\vec{q}|)}}e^{{i\over\hbar}\int^T_{-T}dt
{\em L}(\phi,c,c^+,t)},
\end{equation}
where
\begin{equation}
\label{ldia}
{\em L}(\phi,c,c^+,t)=\sum_{\vec{k}}c^+_{\vec{k}}(t)
\left[i\hbar{\partial\over\partial{t}}-\epsilon(\vec{k})-
\Sigma(\vec{k};[\phi])+i\delta_{\vec{k}}\right]c_{\vec{k}}(t),
\end{equation}
\end{mathletters}
$c_{\vec{k}}(t)=\int{d\omega\over2\pi}e^{-i\omega{t}}c_k$
and $\int{D}cDc^+=\Pi_{k}\int{D}c_kDc_k^+$. Notice that 
although $\psi_k(x;[\phi])$'s are in general time-dependent,
$\Sigma(\vec{k};[\phi])$ is independent of $t$ because of
the independence of $\Sigma(\vec{k};[\phi])$ on $\omega$. 
In the $T\rightarrow\infty$ limit, we may integrate out {\em first}
the $\phi$ field in Eq.\ (\ref{ldia}), obtaining 
\begin{mathletters}
\label{lform}
\begin{equation}
\label{seff}
\lim_{T\rightarrow\infty}TrU(T,-T)=\int{D}cDc^+e^{{i\over\hbar}
\int^T_{-T}dt\left[\sum_{\vec{k}}c^+_{\vec{k}}(t)(i\hbar{\partial\over
\partial{t}}-\epsilon(\vec{k})+i\delta_{\vec{k}})c_{\vec{k}}(t)
-E_{int}[n(\vec{k},t)]\right]},
\end{equation}
where 
\begin{equation}
\label{Eint}
E_{int}[n(\vec{k},t)]
\sim(i\hbar)ln\left[\int{D}\phi{e}^{i\int
^{\infty}_{-\infty}dt'\sum_{\vec{q}}{\phi(\vec{q},t')
\phi(-\vec{q},t')\over2v(|\vec{q}|)}-{i\over\hbar}\sum_{\vec{k}}
\Sigma(\vec{k};[\phi])n(\vec{k},t)}\right],
\end{equation}
\end{mathletters}
where $n(\vec{k},t)=c^+_{\vec{k}}(t)c_{\vec{k}}(t)$.
Notice that we have extented the limit of time integration over $\phi$ fields
from $\pm{T}$ to $\pm\infty$ in Eq.\ (\ref{Eint}). This is valid if $T$ is much
larger than the characteristic time scale $\hbar/E_{\phi}$ 
governing the fluctuations of $\phi$ field, where $E_{\phi}$ is the
corresponding characteristic energy . Notice also that
although the precise form of $E_{int}$ is hard to
determine, it is clear that $E_{int}$ is a functional only of the 
occupation numbers $n(\vec{k})$. The special form of action \
(\ref{seff}) implys that in the long-time, low energy ($<<E_{\phi}$)
limit, the 
Hamltonian of the system, when expressed in terms of the
occupation numbers $n(\vec{k})$'s, has the diagonal form
\[
H=\sum_{\vec{k}}\epsilon(\vec{k})n(\vec{k})+E_{int}[n(\vec{k})], 
\]
i.e., the low energy eigenstates of the system are completely characterised
by occupation numbers $n(\vec{k})$, suggesting that the states 
represented by $n(\vec{k})$ are in fact the quasi-particle states in
Landau Fermi liquid theory\cite{l1,l2}. 

   To show that $n(\vec{k})$ indeed characterizes quasi-particle states
we examine the ground state {\em quasi-particle} occupation
number $<n(\vec{k})>=<n(\vec{k};[\phi])>_{\phi}$, where
\[
n(\vec{k};[\phi])={\hbar\over2\pi{i}}\int_{-\infty}^{\infty}d\omega
{e^{-i\omega0^-}\over\hbar\omega-\epsilon(\vec{k})-\Sigma(\vec{k};
[\phi])+i\delta_{\vec{k}}}=\theta(-\xi_{\vec{k}}),
\]
is independent of $\phi$, and
\begin{equation}
\label{nk2}
<F[\phi]>_{\phi}={\int{D}\phi\left(F[\phi]\right)e^{iS_{eff}(\phi)}\over
\int{D}\phi{e}^{iS_{eff}(\phi)}},
\end{equation}
where $S_{eff}(\phi)\sim\left(\sum_q{\phi(q)\phi(-q)\over2v(|\vec{q}|)}-i
\sum_kln(1+G_0(0,k)\Sigma(\vec{k};[\phi]))\right)$
is obtained from Eq.\ (\ref{lcc}) by first integrating
over the fermion fields. It is easy to see that
$<n(\vec{k})>=\theta(-\xi_{\vec{k}})$, i.e.,
the quasi-particle ground state occupation number is a
$\theta$-function as the occupation number for
non-interacting fermions, in agreement with Fermi liquid theory\cite{l1,l2}.
Notice this is a direct consequence of adiabaticity requirement, that
for any configuration $\phi$, the ground state is obtained by
switching on $\phi$ adiabatically, and the occupation number $n(\vec{k};
[\phi])$ is not affected by the switching process. 
Notice also that in an interacting system, the quasi-particle
wavefunction is a many-body wave-function. This is reflected
in our formalism where $\psi_k$'s are functionals of auxillary
$\phi$ field which represents density fluctuations in the system.
The many-body nature of $\psi_k[\phi]$'s appears as a weighted sum
over all possible configurations of $\phi$ (or density
fluctuations) carried by quasi-particles 
when physical quantities are computed. 

 For small derivations $\delta{n}(\vec{k})=n(\vec{k})-\theta(-\xi_{\vec{k}})$,
the energy of the system is given by the Landau expression,
\begin{equation}
\label{elandau}
E[n(\vec{k})]\sim{E}_0+\sum_{\vec{k}}E_{\vec{k}}\delta{n}(\vec{k})+
{1\over2}\sum_{\vec{k},\vec{k}'}f_{\vec{k}\vec{k}'}\delta{n}(\vec{k})
\delta{n}(\vec{k}'),
\end{equation}
 where $E_0$ is the ground state energy and the quasi-particle
 energy dispersion $E_{\vec{k}}=
\epsilon(\vec{k})+\Sigma(\vec{k})$ and Landau parameters $f_{\vec{k}
\vec{k}'}$ can be computed in our scheme as functional derivatives
of $E_{int}$ with respect to $n(\vec{k})$'s. It is straightforward to show
from Eqs.\ (\ref{lform}) and \ (\ref{nk2}) that $\Sigma(\vec{k})=
\delta{E}_{int}/\delta{n}(\vec{k})=
<\Sigma(\vec{k};[\phi])>_{\phi}$ and $f_{\vec{k}\vec{k}'}=
\delta\Sigma(\vec{k})/\delta{n}(\vec{k'})=
<\Sigma(\vec{k};[\phi]\Sigma(\vec{k'};[\phi])>^c_{\phi}$, where
$<AB>^c_{\phi}=<AB>_{\phi}-<A>_{\phi}<B>_{\phi}$.

   The transport equation for quasi-particles can be constructed 
following the phenomenological approach of Landau\cite{l2} where
{\em local} distribution of quasi-particles
$n(\vec{r},\vec{k};t)$ is introduced and the transport equation
is derived using Eq.\ (\ref{elandau}) by assuming validity of usual classical
kinetic theory. It has to be emphasised that the quasi-particle
states which are exact eigenstates of the system are
extended throughout whole space\cite{pn} and quasi-particles
localized in space are only approximate eigenstates of the system and
have finite life-time $\tau$\cite{l2,pn}. To derive the 
transport equation more rigorously we have to define first local
quasi-particle states in our formulation. An effective low energy
Hamiltonian for the {\em local} quasi-particles can then be derived
to obtain the correct transport equation. 

   For non-interacting system a localized state can be formed by 
superposition of plane wave states. However, in our formulation
since quasi-particle wavefunctions are functionals of $\phi$
which characterize density fluctuations in the system, a local
quasi-particle state can be constructed only if we restrict
also density fluctuations (or $\phi$) carried by the quasi-particle
to be local also. For {\em local} quasi-particle state of size
scale $q^{-1}$, we expect that the $\phi$ field associated with
the quasi-particle should be restricted to momentum scale $>q_c\sim
{q}$. To implement this scheme we divide the $\phi$ field into
slow and fast varying parts $\phi(x)=\phi_s(x)+\phi_f(x)$, where
$\phi_{s(f)}(x)=\sum_{|\vec{q}|<(>)q_c}e^{i\vec{q}.\vec{x}}\phi(q)$
, and we shall assume
that local quasi-particles of size $q_c$ can be constructed from
superposition of quasi-particle states of the corresponding 
interacting system, when $\phi(x)$ is replace by $\phi_f(x)$.
Notice that for a system of size $L\sim{q}_c^{-1}$, our local
quasi-particle states become exact eigenstates of the system, as is 
expected physically. 

  With this definition of {\em local} quasi-particles we can derive
the effective low energy Hamiltonian and transport equation
for local quasi-particles straightforwardly. The algebra 
of this derivation is lengthy and we shall
report the details in a seperate paper. We find that
the usual Fermi liquid transport equation is recovered in the limit
$q_c\rightarrow0$\cite{ng}.

   Lastly we examine the quasi-particle renormalization factor 
{\em z}. To compute {\em z}, we examine the usual one-particle
Green's function $g(\vec{k},\omega)$ in our theory. Using Eq.(5),
it is straightforward to show that
\begin{equation}
\label{g1}
g(\vec{k},\omega)=\sum_{k'}<\left({|<\psi_k^{(0)}|\psi_{k'}>|^2\over
\hbar\omega'-\epsilon(\vec{k}')-\Sigma(\vec{k}';[\phi])+i\delta_{\vec{k}'}}
\right)>_{\phi},
\end{equation}
where $k'=(\vec{k}',\omega')$ and $<\psi_k^{(0)}|\psi_{k'}>=\int{d}^{d+1}
x\psi_k^{(0)*}(x)\psi_{k'}(x;[\phi])$. The bare-particle occupation
number is 
\[
n^b(\vec{k})={\hbar\over2\pi{i}}\int{d}\omega{g}(\vec{k},\omega)
=\sum_{\vec{k'}}\theta(-\xi_{\vec{k}})<|F_{\vec{k}\vec{k}'}
^{\phi}(t)|^2>_{\phi}, \]
where $F_{\vec{k}\vec{k}'}^{\phi}(t)=
\int{d^dx\over(2\pi)^d}e^{-i\vec{k}.\vec{x}}\psi_{\vec{k}',0}(\vec{x},t;
[\phi])$. We have again made used of the result $\psi_{\vec{k},\omega+\Omega}
(x;[\phi])=e^{-i\Omega{t}}\psi_{\vec{k},\omega}(x;[\phi])$ in deriving
the last expression. Assuming that the wavefunctions
$\psi_{k}(x;[\phi])$ are smooth functions of $\vec{k}$ which
do not show any discontinuity across Fermi surface, it is easy to show
using Eq.\ (\ref{ls}) that, $z=
n^b(k_F^+)-n^b(k_F^-)=<|A_{k_F}[\phi]|^2>_{\phi}$,
i.e., $z$ is equal to the weighted average over $\phi$ field the
wavefunction overlap between non-interacting states
$\psi_k^{(0)}(x)$ and $\psi_k(x;[\phi])$ on the Fermi surface,
as is expected from our identification of $\psi_k(x;[\phi])$'s as
quasi-particles wavefunctionals in our theory.

  A few comments on our formulation of Fermi liquid theory is now in
order. By organizing perturbation theory in a slightly different way, 
we have succeeded in formulating Fermi liquid theory directly 
in terms of quasi-particles in our theory. 
Notice that as far as an order-by-order expansion in computing
quantities like one- and two- particle Green's functions are 
concerned, our formulation offers nothing new. However, the possibility
of identifying quasi-particles in our theory enables us to compute
quantities like Landau-parameters and quasi-particle energies more
directly. For example, it is straightforward to
show that by keeping the self-energy $\Sigma(\vec{k};[\phi])$
to second order in $\phi$, we obtain the usual RPA (Random-Phase
Approximation) expression for ground state energy, with the occupation
number $n^{(0)}(\vec{k})$ in the Lindhard function $\chi_0$ replaced
by the corresponding quasi-particle occupation number 
in our theory. The quasi-particle self-energy $\Sigma(\vec{k})$
is found to be equal to the RPA on-shell self-energy
$\Sigma^{RPA}(\vec{k},\xi_{\vec{k}})$. Notice that in general there
are more than one ways of introducing the Hubbard-Strotonovich
transformation and the auxillary fields, corresponding
to different ways of organizing perturbation series\cite{no2}.
Our results remain unchanged as long as
single-particle wave-functionals $\psi_k(x;[\phi])$
can be introduced and can be obtained perturbatively from the
non-interacting states $\psi_k^{(0)}(x)$. 

   The main power of our formulation appears in studying Fermi liquids 
with $z\rightarrow0$, where usual perturbation theory fails. Notice
that $z\rightarrow0$ does not necessarily imply breakdown of adiabaticity.
For example, for a system of volume $L^d$, and in the presence of 
auxillary field $\phi$, an eigenstate $\psi_k$ is formed by
mixing plane wave state $\psi_k^{(0)}$ with other
plane wave states $\psi_{k'}^{(0)}$. The probability of mixing
is of order $1/L^d$ for each state $\vec{k'}\neq\vec{k}$. As
long as the renormalization factor $A_{k}[\phi]$ is of order
$\geq{1/L^d}$, the state $\psi_k$ may still 'remembers' its parent
plane wave state $\psi_{k}^{(0)}$ and adiabaticity can be kept,
although $z$ may approach zero in the $L\rightarrow\infty$ limit. 
Our formulation implys that a Fermi-Liquid description may still be
applicable in describing the thermodynamics and low-energy,
long-wavelength propteries of these systems where the one-particle 
properties may be drastically different from usual $z\neq0$
Fermi liquids. Examples of these systems include
one-dimensional Luttinger liquids\cite{hl} where similar idea
has been proposed by Carmelo {\em et.al.}\cite{ca} and systems with
super-long-range interactions in dimensions $>1$\cite{wen}. We
find that our Fermi-liquid formulation, after suitably refined,
can be applied to describe these systems\cite{ng} where the
one-particle properties are non-Fermi liquid like. 
The details of our calculations will be
presented in a seperate paper.


\end{document}